\documentclass[12pt,aps]{revtex4}%
\usepackage{amssymb}
\usepackage{amsmath}
\usepackage{amsfonts}
\usepackage{graphicx}%
\providecommand{\U}[1]{\protect\rule{.1in}{.1in}}
\providecommand{\U}[1]{\protect\rule{.1in}{.1in}}
\providecommand{\U}[1]{\protect\rule{.1in}{.1in}}
\providecommand{\U}[1]{\protect\rule{.1in}{.1in}}
\providecommand{\U}[1]{\protect\rule{.1in}{.1in}}
\providecommand{\U}[1]{\protect\rule{.1in}{.1in}}
\providecommand{\U}[1]{\protect\rule{.1in}{.1in}}
\providecommand{\U}[1]{\protect\rule{.1in}{.1in}}
\providecommand{\U}[1]{\protect\rule{.1in}{.1in}}
\providecommand{\U}[1]{\protect\rule{.1in}{.1in}}
\providecommand{\U}[1]{\protect\rule{.1in}{.1in}}
\providecommand{\U}[1]{\protect\rule{.1in}{.1in}}
\providecommand{\U}[1]{\protect\rule{.1in}{.1in}}
\providecommand{\U}[1]{\protect\rule{.1in}{.1in}}
\providecommand{\U}[1]{\protect\rule{.1in}{.1in}}
\providecommand{\U}[1]{\protect\rule{.1in}{.1in}}
\providecommand{\U}[1]{\protect\rule{.1in}{.1in}}
\providecommand{\U}[1]{\protect\rule{.1in}{.1in}}
\begin{document}
\title{Killing Vectors and Anisotropy}
\author{J.P. Krisch and E.N. Glass}
\affiliation{Department of Physics, University of Michigan, Ann Arbor, MI 48109}
\date{14 May 2009}

\begin{abstract}
We consider an action that can generate fluids with three unequal stresses for
metrics with a spacelike Killing vector. The parameters in the action are
directly related to the stress anisotropies. \ The field equations following
from the action are applied to an anisotropic cosmological expansion and an
extension of the Gott-Hiscock cosmic string.

\end{abstract}
\maketitle

\section{Introduction}

General relativistic models with anisotropic stress have become increasing
useful as the applications become more complex and precise \cite{HOP08, HSW08,
CH08, ACM08, KG08, KM08}.\ There has been considerable interest in anisotropic
spheres \cite{BL74, CHE+81, Bay82, Bon92,Via08} because of applications to
stellar models \cite{DG02, MH03, DG04}, and temperature anomalies in the CMB
\cite{OTZ+04,RS07} have generated an increased interest in anisotropic
cosmological models \cite{CH08, GCP07, PPU07, PPU08, DD07, MKK+07}. Anisotropy
is usually discussed in the context of fluid stress-energy relations, and
examining the Lagrangian actions that generate fluid equations of state
provides a simple way to view their geometric and physical origins.\ This
paper discusses a simple action describing a fluid with three unequal
stresses.\ The fluid is supported by a metric with a spacelike Killing
symmetry. \ 

The vacuum Einstein field equations, for spacetimes with Killing vector
$\xi^{a}$, can be generated from an action written in terms of the Killing
vector norm $\lambda=\xi^{a}\xi_{a},$ and twist $\omega_{i}=\varepsilon
_{iabc}\xi^{a}\nabla^{b}\xi^{c}$.\ The original vacuum action was developed by
Ehlers \cite{Ehl62}, Harrison \cite{Har68}, and Geroch \cite{Ger71}, who wrote
the 3+1 Einstein equations\ on the 2+1 space of Killing vector orbits. For
spacetimes with metric $g_{ab,}$ $(-+++),$ the induced metric on the
3-manifold of Killing orbits is $h_{ab}=g_{ab}-\xi_{a}\xi_{b}/\lambda$. The
action generating the vacuum field equations is, with $h_{ab}$ replaced by
$\gamma_{ab}:=\lambda h_{ab}$,\
\begin{equation}
S_{vac}=\int d^{3}x\sqrt{\gamma}(\mathcal{R}-\frac{D_{a}\lambda D^{a}%
\lambda+D_{a}\omega D^{a}\omega}{2\lambda^{2}})
\end{equation}
where $\mathcal{R}$ and $D_{a}$ are the Ricci scalar and covariant derivative
on $\gamma$ respectively, and $\omega_{a}=D_{a}\omega.$ This action was
generalized to perfect fluids \cite{Ste88,GGK97} by including a scalar
function $\mathcal{K}$, and a vector function, $s^{a}$ with $\xi^{a}s_{a}=0.$
\
\begin{equation}
S_{fluid}=\int d^{3}x\sqrt{\gamma}(\mathcal{R}-\frac{D_{a}\lambda D^{a}%
\lambda+D_{a}\omega D^{a}\omega}{2\lambda^{2}}-\mathcal{K}-s^{a}s_{a})
\end{equation}
Using this action, a stress-energy for two isotropic equations of state can be
generated.\ Krisch and Glass \cite{KG02} showed that the same action may also
be applied to fluids with two unequal stresses. \ 

In this paper, which considers metrics with a spacelike Killing vector, the
action $S_{fluid}$ is extended to completely anisotropic fluids. Previous work
using this action treated $\mathcal{K}$ and $s^{a}s_{a}$ separately.
Considering both terms simultaneously allows extensions to fluids with three
unequal stresses. In the next section we develop the anisotropic fluid content
implicit in $S_{fluid}$, and discuss some of the equations of state and their
effect on the rate-of-expansion of the fluid 4-velocity.\ Some particular
metric examples are given in the third section. \ 

\section{Fluid Content}

\subsection{Field Equations and Geometry}

The 2+1 field equations that follow from the action $S_{fluid}$ are
\begin{align}
\mathcal{R}_{ab}=\frac{D_{a}\lambda D_{b}\lambda+D_{a}\omega D_{b}\omega
}{2\lambda^{2}}+s_{a}s_{b}+\mathcal{K}\gamma_{ab}  & \label{ricci-1}\\
\lambda^{2}D^{a}(\lambda^{-1}D_{a}\lambda)+D_{a}\omega D^{a}\omega=0  &
\label{pot-1}\\
D_{a}[\frac{D^{a}\omega}{\lambda^{2}}]=0  &
\end{align}
where $D_{a}$ is the covariant derivative on $\gamma.$ $\mathcal{R}_{ab}=$
Ricci($\gamma_{ab}$) is related to $R_{ab}=$ Ricci($g_{ab}$) \cite{Ger71,
GGK97} by
\begin{align}
h_{a}^{\ c}h_{b}^{\ d}R_{cd}  &  =\mathcal{R}_{ab}-\frac{D_{a}\lambda
D_{b}\lambda+D_{a}\omega D_{b}\omega}{2\lambda^{2}}=s_{a}s_{b}%
+\mathcal{K\lambda}h_{ab}\label{ssk-term}\\
\xi^{a}\xi^{b}R_{ab}  &  =0\\
h_{a}^{\ c}\xi^{b}R_{cb}  &  =0
\end{align}
In the next section, we discuss the stress-energy related to the matter
parameters, $s_{a}$ and $\mathcal{K}$.

\subsection{Stress-Energy}

A simple anisotropic fluid description contains a density and three stresses
$(\varepsilon,P_{1},P_{2},P_{3})$ which depend on $s_{a}$ and $\mathcal{K}.$
In this section we find the dependence of each of the stress-energy components
on the action parameters and show that the spatial part of $s_{a}$ is a
measure of the stress anisotropy $P_{1}-P_{2}$, while $\mathcal{K}$ enters
into the anisotropy between the $(1,3)$ and $(2,3)$ planes. Consider a 3+1
metric, $g_{ab}$, described with unit tetrad vectors $[U^{a},e_{(1)}%
^{a},e_{(2)}^{a},e_{(3)}^{a}]$, and with spacelike Killing vector $\xi^{a}$
aligned with $e_{(3)}^{a}$: $\xi^{a}=\sqrt{\lambda}e_{(3)}^{a}.$ The general
stress-energy tensor is assumed to be
\begin{equation}
T_{ab}=\varepsilon U_{a}U_{b}+P_{1}e_{a}^{(1)}e_{b}^{(1)}+P_{2}e_{a}%
^{(2)}e_{b}^{(2)}+P_{3}e_{a}^{(3)}e_{b}^{(3)} \label{gen-en-mom}%
\end{equation}
with the stress components carrying a tetrad index. The Ricci tensor for this
stress-energy is
\begin{align*}
R_{ab}  &  =8\pi{\LARGE [}\frac{\varepsilon+P_{1}+P_{2}+P_{3}}{2}U_{a}%
U_{b}+\frac{\varepsilon+P_{1}-P_{2}-P_{3}}{2}e_{a}^{(1)}e_{b}^{(1)}\\
&  +\frac{\varepsilon-P_{1}+P_{2}-P_{3}}{2}e_{a}^{(2)}e_{b}^{(2)}%
+\frac{\varepsilon-P_{1}-P_{2}+P_{3}}{2}e_{a}^{(3)}e_{b}^{(3)}{\LARGE ]}.
\end{align*}
Using the field equation $\xi^{b}\xi^{a}R_{ab}=0,$ $P_{3}$ is determined by
the other stress-energy components%
\begin{equation}
P_{3}=-\varepsilon+P_{1}+P_{2}%
\end{equation}
and Ricci tensor becomes
\begin{equation}
R_{ab}=8\pi\lbrack(P_{1}+P_{2})U_{a}U_{b}+(\varepsilon-P_{2})e_{a}^{(1)}%
e_{b}^{(1)}+(\varepsilon-P_{1})e_{a}^{(2)}e_{b}^{(2)}].
\end{equation}
Expanding $\ s^{a}$, using $R_{ab}=s_{a}s_{b}+\lambda\mathcal{K}h_{ab}$ and
taking scalar products, the complete stress energy description is
\begin{subequations}
\label{p-s-eq}%
\begin{align}
s_{a}  &  =s_{0}U_{a}+s_{1}e_{a}^{(1)}+s_{2}e_{a}^{(2)}\label{ps-a}\\
16\pi\varepsilon &  =(s_{0})^{2}+(s_{1})^{2}+(s_{2})^{2}+\lambda
\mathcal{K}\label{ps-b}\\
16\pi P_{1}  &  =(s_{0})^{2}+(s_{1})^{2}-(s_{2})^{2}-\lambda\mathcal{K}%
\label{ps-c}\\
16\pi P_{2}  &  =(s_{0})^{2}-(s_{1})^{2}+(s_{2})^{2}-\lambda\mathcal{K}%
\label{ps-d}\\
16\pi P_{3}  &  =(s_{0})^{2}-(s_{1})^{2}-(s_{2})^{2}-3\lambda\mathcal{K}%
\label{ps-e}\\
s_{1}s_{2}  &  =s_{0}s_{1}=s_{0}s_{2}=0 \label{ps-f}%
\end{align}
where $s_{0}$, $s_{1},$ and $s_{2}$ are tetrad indexed.\ These relations
between the stress-energy and the action parameters show how $s_{a}$ and
$\mathcal{K}$ enter the fluid anisotropy.\ Using Eq.(\ref{ps-c}) and
Eq.(\ref{ps-d}), the anisotropy in the 2+1 stress is described by the spatial
components of $s_{a}$:
\end{subequations}
\begin{equation}
8\pi(P_{1}-P_{2})=(s_{1})^{2}-(s_{2})^{2}. \label{pdif}%
\end{equation}
The anisotropies involving the stress along the Killing direction require both
$s_{a}$ and $\mathcal{K}$.
\begin{align}
8\pi(P_{1}-P_{3})  &  =s_{1}^{2}+\lambda\mathcal{K},\\
8\pi(P_{2}-P_{3})  &  =s_{2}^{2}+\lambda\mathcal{K}\nonumber
\end{align}

A particularly useful relation emerging from this formalism is the equality of
the stress along the Killing direction with the negative Ricci scalar: $16\pi
P_{3}=-R$.\ 

\subsection{Equations of State}

Some equations of state for the 3+1 fluid can be written down by considering
values for $s_{a}$ and $\mathcal{K}$. The conditions in Eq.(\ref{ps-f}) mean
that any one of $s_{0}$, $s_{1}$, or $s_{2}$ can be non-zero. In the formalism
so far, the choice $s_{1}$ or $s_{2}$ equal to zero only fixes the $(1,2)$
index. The stress-energy relations for each of the choices are described in
Table I\ (with $s_{2}=0$) and $s_{1}$ identified as the index function for
stress anisotropy in the $(1,2)$ plane. The $s_{1}=0$ choice simply replaces
$2$ by $1$ in Table 1.

\begin{center}
Table 1
\[%
\begin{tabular}
[c]{|l|l|}\hline
non-zero functions & $\ \ \ \ \ \ \ \text{stress-energy relations
\ \ \ \ \ \ \ \ \ \ \ }$\\\hline
$\mathcal{K}\text{\ \ \ }$ & $\varepsilon=-P_{1}=-P_{2}=-P_{3}/3$\\\hline
$s_{1}$ & $\varepsilon=P_{1}=-P_{2}=-P_{3}$\\\hline
$s_{0}$ & $\varepsilon=P_{1}=P_{2}=P_{3}$\\\hline
$\mathcal{K},$ $s_{1}$ & $\varepsilon=-P_{2},$ $P_{3}=2P_{2}+P_{1}$\\\hline
$\mathcal{K}$, $s_{0}$ & $P_{1}=P_{2},$ $P_{3}=-\varepsilon+2P_{1}$\\\hline
\end{tabular}
\ \ \ \
\]

\end{center}

The fluid parameter conditions can be generally related to the value of
$s^{a}s_{a}$. For $s_{2}=0$, by using Eq.(\ref{p-s-eq}) we find%
\begin{equation}
8\pi(\varepsilon+2P_{2}-P_{1})=-s^{a}s_{a}.
\end{equation}
If $s_{a}$ is timelike, $s_{1}=0$, $\varepsilon+2P_{2}>P_{1}$, and
$P_{1}=P_{2}$. The timelike condition is $\varepsilon+P_{1}>0,$ $P_{1}%
=P_{2}.\ $If $s_{a}$ is spacelike\ with $s_{2}=0$, we have $\varepsilon
+2P_{2}<P_{1}$. Since only one component of $s_{a}$ is non zero, $s_{a}$
cannot be null and we do not have $\varepsilon+2P_{2}=P_{1\text{ }}.$
Completely anisotropic fluids will correspond to $s_{1}$ (or $s_{2}$) $\neq0,$
$\mathcal{K}\neq0.$ The fluid conditions for a completely anisotropic fluid
are%
\begin{subequations}
\begin{align}
s_{2}  &  =0:\varepsilon=-P_{2},\text{\ \ }P_{3}=2P_{2}+P_{1}\\
s_{1}  &  =0:\varepsilon=-P_{1},\text{\ \ }P_{3}=2P_{1}+P_{2}%
\end{align}

\subsection{Energy Conditions}

There are a number of energy conditions based on the structure of the
stress-energy and Ricci tensor \cite{Poi04}. In matter, the weak and dominant
conditions require positive density and $-T_{ab}U^{b}=\varepsilon U_{a}.$ With
either $s_{1}$ or $s_{2}$ non zero and a comoving observer, the strong and
null conditions are
\end{subequations}
\begin{align}
\text{Strong}  &  :R_{ab}U^{a}U^{b}\geq0\\
-\lambda\mathcal{K}  &  \mathcal{=}8\pi(P_{1}+P_{2})\geq0\\
\text{Null}  &  :T_{ab}N^{a}N^{b}\geq0\\
1.\text{ }N^{a}  &  =U^{a}+e_{(1)}^{a}:\varepsilon+P_{1}\geq0\Rightarrow
s_{1}^{2}\geq0\nonumber\\
2.\text{ }N^{a}  &  =U^{a}+e_{(2)}^{a}:\varepsilon+P_{2}\geq0\Rightarrow
s_{2}^{2}\geq0\nonumber\\
3.\text{ }N^{a}  &  =U^{a}+e_{(3)}^{a}:\varepsilon+P_{3}\geq0\Rightarrow
(P_{1}+P_{2})\geq0\nonumber
\end{align}
The weak and dominant conditions can be enforced, but the strong and some of
the null conditions can be violated. For example, the $(s_{0},s_{2}=0)$
condition, $\varepsilon=-P_{2},$ implies a violation if $\left\vert
P_{2}\right\vert >P_{1}$.\ The violation of the strong energy condition is
related to the development of the 4-velocity rate-of-expansion, $\Theta
=\nabla_{a}U^{a}$, in the Raychaudhuri equation%
\begin{equation}
\frac{d\Theta}{d\tau}=-\frac{1}{3}\Theta^{2}-\sigma^{ab}\sigma_{ab}%
+\omega^{ab}\omega_{ab}-R_{ab}U^{a}U^{b}.
\end{equation}
The action considered in this paper determines the form of the Ricci tensor.
For the completely anisotropic fluid, the Ricci tensor is
\begin{align}
s_{2}  &  =0:R_{ab}=8\pi\lbrack-(P_{1}+P_{2})(g_{ab}-e_{a}^{(3)}e_{b}%
^{(3)})+(P_{1}-P_{2})e_{a}^{(1)}e_{b}^{(1)}]\\
&  =\lambda\mathcal{K}(g_{ab}-e_{a}^{(3)}e_{b}^{(3)})+s_{1}^{2}e_{a}%
^{(1)}e_{b}^{(1)}\nonumber\\
s_{1}  &  =0:R_{ab}=8\pi\lbrack-(P_{1}+P_{2})(g_{ab}-e_{a}^{(3)}e_{b}%
^{(3)})+(P_{2}-P_{1})e_{a}^{(2)}e_{b}^{(2)}]\\
&  =\lambda\mathcal{K}(g_{ab}-e_{a}^{(3)}e_{b}^{(3)})+s_{2}^{2}e_{a}%
^{(2)}e_{b}^{(2)}\nonumber
\end{align}
The Raychaudhuri equation becomes%
\begin{align}
\frac{d\Theta}{d\tau}  &  =-\frac{1}{3}\Theta^{2}-\sigma^{ab}\sigma
_{ab}+\omega^{ab}\omega_{ab}+\lambda\mathcal{K},\\
\lambda\mathcal{K}  &  =-8\pi(P_{1}+P_{2}).\nonumber
\end{align}
With zero vorticity and $\lambda\mathcal{K}=0,$ $\dot{\Theta}<0$, these
conditions describe a decreasing rate-of-expansion with an initially
converging or diverging time-like congruence becoming more focused.\ However,
even for zero vorticity, if $\ \lambda\mathcal{K}>$ $0$, the rate-of-change of
the expansion can be positive, $\dot{\Theta}>0$, and is unfocusing.

\subsection{Fluid Shear and Anisotropy}

The form of the stress-energy tensor, Eq.(\ref{gen-en-mom}), identifies all of
the spatial components as stress. However, equivalences \cite{KE73, Tup81} can
be used to relate the completely anisotropic form to fluids with shear, with
the differences in the shear tensor identified as the physical cause of the
anisotropy.\ The general equivalence relation for $s_{1}\neq0$ is
\begin{align}
\varepsilon U_{a}U_{b}+P_{1}e_{a}^{(1)}e_{b}^{(1)}+P_{2}e_{a}^{(2)}e_{b}%
^{(2)}+P_{3}e_{a}^{(3)}e_{b}^{(3)}  &  =(\varepsilon+P)U_{a}U_{b}%
+Pg_{ab}-2\eta\sigma_{ab}\\
\varepsilon=-P_{2}  & \nonumber\\
P_{1}=P-2\eta\sigma_{(11)}  & \nonumber\\
P_{2}=P-2\eta\sigma_{(22)}  & \nonumber\\
P_{3}=P-2\eta\sigma_{(33)}  & \nonumber
\end{align}
The trace-free condition for $\sigma_{ab}$ provides $P=(P_{1}+P_{2}+P_{3})/3.$
Using Eq.(\ref{p-s-eq}) the action parameters $s^{a}$ and $\mathcal{K}$
describe the differences in the shear tensor components. \
\begin{align}
P_{i}  &  =P-2\eta\sigma_{(ii)},\\
s_{1}^{2}  &  =16\pi\eta\lbrack\sigma_{(22)}-\sigma_{(11)}],\nonumber\\
\lambda\mathcal{K}  &  =16\pi\eta\lbrack\sigma_{(33)}-\sigma_{(22)}].\nonumber
\end{align}
Imposing an extra condition identifies an equation of state and the tetrad
indexed shear tensor components. The evolution of spherically symmetric fluids
of this type has been considered by Herrera et al \cite{HSW08}, who gave a
metric for a shearing expansion-free evolving fluid with two unequal stresses.

\section{Examples}

The stress-energy of a completely anisotropic fluid is described by\
\begin{subequations}
\begin{align}
s_{2}  &  =0:\varepsilon=-P_{2},\text{ \ \ }P_{3}=2P_{2}+P_{1}\\
s_{1}  &  =0:\varepsilon=-P_{1},\text{ \ \ }P_{3}=2P_{1}+P_{2}%
\end{align}
$P_{2}$ $(P_{1})$ must be negative for positive density. This suggests two
possible application areas:\ Bianchi cosmological models, and cosmic strings
with an axial tension.\ \ 

\subsection{Anisotropic Cosmology}

Bianchi metrics have been used to model anomalies in the CMB radiation
\cite{GCP07, PPU07, KM08, CH08, PPU08}, with anisotropies developing due to
different expansion rates along coordinate axes.\ A general metric to consider
is
\end{subequations}
\begin{equation}
ds^{2}=-dt^{2}+b^{2}(t)dx_{1}^{2}+c^{2}(t)dx_{2}^{2}+f^{2}(t)dx_{3}^{2}%
\end{equation}
with three Hubble rates, $H_{1}=\dot{b}/b,$ $H_{2}=\dot{c}/c,$ and\ $H_{3}%
=\dot{f}/f$.\ For a power law expansion, one has $b=b_{0}t^{\beta},$
$c=c_{0}t^{\gamma},$ $f=f_{0}t^{\delta}$. The stress-energy is%
\begin{align}
8\pi\varepsilon(t^{2})  &  =\beta\gamma+\beta\delta+\gamma\delta
\label{pl-stress-energy}\\
8\pi P_{1}(t^{2})  &  =-\gamma(\gamma-1)-\delta(\delta-1)-\delta
\gamma\nonumber\\
8\pi P_{2}(t^{2})  &  =-\beta(\beta-1)-\delta(\delta-1)-\beta\delta\nonumber\\
8\pi P_{3}(t^{2})  &  =-\beta(\beta-1)-\gamma(\gamma-1)-\gamma\beta\nonumber
\end{align}
The stress-energy conditions $\varepsilon=-P_{2}$ and $P_{3}=2P_{2}+P_{1}$
impose constraints%
\begin{align}
\gamma(\beta+\delta)  &  =\beta(\beta-1)+\delta(\delta-1)\label{const-1}\\
\gamma\beta &  =\beta(\beta-1)+3\delta(\delta-1)+\delta(2\beta+\gamma).
\label{const-2}%
\end{align}
Combining the constraints yields%
\begin{equation}
\delta(\delta-1+\beta+\gamma)=0.
\end{equation}
Two cases emerge: $\delta=0$ and $\delta=1-\beta-\gamma.$ The $\delta\neq0$
condition, when substituted back into Eq.(\ref{const-1}), relates $\beta$ and
$\gamma$%
\begin{equation}
\gamma+\beta=\beta^{2}+\beta\gamma+\gamma^{2}%
\end{equation}
and results in zero energy-density. Only the $\delta=0$ case, with no
expansion along the Killing direction, provides a fluid with non-zero
density.\ With $\delta=0,$ Eq.(\ref{const-1}) is
\begin{equation}
\beta(\beta-1-\gamma)=0.
\end{equation}
When $\beta=0$ the density is zero. In order to have an anisotropic fluid with
positive density, expansions along at least two of the axes are necessary,
with $b(t)=b_{0}t^{\beta}$ and $c(t)=c_{0}t^{\beta-1}$.\ For $\beta\neq0,$ the
stress energy is
\begin{align}
8\pi\varepsilon &  =\beta(\beta-1)/t^{2}\\
8\pi P_{1}  &  =-(\beta-1)(\beta-2)/t^{2}\nonumber\\
8\pi P_{2}  &  =-\beta(\beta-1)/t^{2}\nonumber\\
8\pi P_{3}  &  =-(\beta-1)(3\beta-2)/t^{2}\nonumber\\
\mathcal{K}  &  =2(\beta-1)^{2}/t^{2}\nonumber
\end{align}
This example has positive density for $\beta<0\ $and $\beta>1$, with $\beta>1$
describing expansion.\ The rate-of-expansion for this case is%
\begin{equation}
\Theta=\frac{\dot{b}}{b}+\frac{\dot{c}}{c}=\frac{(2\beta-1)}{t},\text{ \ }%
\dot{\Theta}=\frac{(1-2\beta)}{t^{2}}.
\end{equation}
The congruence will unfocus for $\beta<1/2.$ For the case of an expanding
space with positive density the congruence is focusing.\ The family of radial
stresses parameterized by $\beta$ is specially interesting.\ For $1<\beta<2,$
the stress is a pressure, for $\beta=2$ it is dust, and for $\beta>2$ it is a
tension. $P_{3}$ also shows this range of behavior with the $P_{3}=0$ dust
crossover at $\beta=2/3,$ out of the physical density region.\ Since $P_{3}$
is proportional to the negative Ricci scalar, the behavior of $P_{3}$ also
reflects a change from positive scalar curvature, through zero, to a negative
curvature manifold.\ In this model, only manifolds with non zero Ricci scalar
will have positive density.\ The timelike congruence will unfocus for
$\beta<0.$ This is a positive density region of the parameter space with all
three stresses as tensions.
\[
\text{\textbf{In summary} \ }\beta<0\text{ unfocuses, \ }\beta>1\text{
focuses.}%
\]

\subsection{Cosmic String}

\subsubsection{Metric and Stress-Energy}

Cosmic strings are of current interest both experimentally
\cite{KSV08,FRS+08,KW08,SSS+08} and theoretically \cite{Vac08,CMM+08,
DFM+07}.\ The usual cosmic string models have only a single axial tension. The
simplest family of static 3+1 strings is described by the metric with
$(0,1,2,3)=(t,r,\varphi,z)$%
\begin{equation}
ds^{2}=-dt^{2}+dr^{2}+E^{2}(r)d\varphi^{2}+dz^{2}.
\end{equation}
The density is $8\pi\varepsilon=-E^{\prime\prime}/E,$ and the stresses are%
\begin{equation}
P_{r}=P_{\varphi}=0,\text{ \ }P_{z}=E^{\prime\prime}/E.
\end{equation}
The Gott-Hiscock(GH) static string \cite{Got85,His85} is the constant density
example with $a=\sqrt{8\pi\varepsilon},$ $E(r)=c_{1}\sin(ar)$.\ The matter
metric matches directly to vacuum Levi-Civita with an angular deficit and is
Minkowski near the axis.\ There are two spacelike Killing vectors,
$\xi_{(z)\text{ }}$ and $\xi_{(\phi)}.$

As an anisotropic extension to the\ GH string interior, consider the metric%
\begin{equation}
ds^{2}=-A_{0}^{2}\cos^{2}[a(R_{0}-r)]dt^{2}+dr^{2}+a^{-2}\sin^{2}%
(ar+\gamma)d\phi^{2}+dz^{2}.
\end{equation}
The $\gamma$ contribution to the sine argument is included for finite stress
at $r=0$. The stress-energy for this metric is
\begin{align}
8\pi\varepsilon &  =a^{2}=-8\pi P_{\phi}\\
8\pi P_{r}  &  =a^{2}\tan[a(R_{0}-r)]\cot(ar+\gamma)\nonumber\\
8\pi P_{z}  &  =-a^{2}\{2-\tan[a(R_{0}-r)]\cot(ar+\gamma)\}\nonumber
\end{align}
with $\sin(\gamma)\neq0.$ The axial behavior of this stress-energy is
especially interesting. In the cosmological example, several of the stress
components ranged through pressure, dust, and tension as the metric parameter
varied. In this example, the same behavior is observed but is linked to the
radial position or, along the axis, to the size of the string.\ To interpret
this example as an anisotropic cosmic string with physical tension along the
axis requires $1<\tan[aR_{0}]\cot(\gamma)<2.$ The actual axial structure is
related to the vacuum match and associated angular deficit. These are
discussed in the next sections.

\subsubsection{Vacuum Matching}

The metric can be matched to a vacuum metric across an Israel layer at
$r=R_{0}$.\ The exterior (+) metric is a vacuum Levi-Civita metric with
angular deficit $\delta$ \
\begin{equation}
ds_{\text{LeviCivita}}^{2}=-dt^{2}+dr^{2}+r^{2}\delta^{2}d\phi^{2}+dz^{2}.
\end{equation}
The metric of the Israel layer is%
\begin{equation}
ds_{\text{layer}}^{2}=-A_{0}^{2}dt^{2}+a^{-2}\sin^{2}(aR_{0}+\gamma)d\phi
^{2}+dz^{2}.
\end{equation}
The Levi-Civita metric is matched to the layer. The matching conditions are
\begin{equation}
A_{0}=1,\text{ \ \ \ }a^{-1}\sin(aR_{0}+\gamma)=\pm R_{0}\delta.
\label{layer-match}%
\end{equation}
The stress-energy of the Israel boundary layer is calculated from jumps in the
extrinsic curvature, $K_{ab}$, going from Levi-Civita $(+)$ across the Israel
layer to the string interior $(-)$, with the jumps calculated from
$<K_{ab}>\ =K_{ab}^{+}-K_{ab}^{-},$ $K=K_{a}^{a}.$ The layer stress-energy is
\begin{equation}
-8\pi S_{ab}=\ <K_{ab}>-<K>g_{ab}^{\text{layer}}.
\end{equation}
The extrinsic curvatures needed to calculate the jumps are
\begin{align}
K_{\phi\phi}^{+}  &  =-\delta^{2}R_{0},\text{ \ \ \ }K_{tt}^{+}=0,\text{
\ \ \ }K_{zz}^{+}=0,\text{ }\\
K_{\phi\phi}^{-}  &  =a^{-1}\sin(aR_{0}+\gamma)\cos(aR_{0}+\gamma),\text{
\ \ }K_{tt}^{-}=0,\text{ \ \ }K_{zz}^{-}=0.\nonumber
\end{align}
The surface of the anisotropic string has an Israel stress-energy content%
\begin{align}
8\pi S_{tt}  &  =-<K>,\text{ \ }8\pi S_{\phi\phi}=0,\text{ \ \ }8\pi
S_{zz}=\ <K>\\
&  <K>\ =-1/R_{0}-a\cot(aR_{0}+\gamma)\ \nonumber
\end{align}
The equation of state of the GH string solution is found in the boundary layer
of the anisotropic string.

\subsubsection{Angular Deficit}

In the GH static string, the angular deficit is related to the mass/length,
$\mu,$ calculated from a $t=const$, $z=const$ integral of the density. For the
anisotropic string, that mass is composed of two parts.\ The contribution from
the string interior is%
\begin{align}
\mu_{1}  &  =%
{\displaystyle\int\limits_{0}^{2\pi}}
{\displaystyle\int\limits_{0}^{R_{0}}}
\varepsilon\frac{\sin(ar+\gamma)}{a}drd\phi=2\pi%
{\displaystyle\int\limits_{0}^{R_{0}}}
\frac{a^{2}}{8\pi}\frac{\sin(ar+\gamma)}{a}dr\\
&  =\frac{1}{4}[\cos(\gamma)-\cos(aR_{0}+\gamma)],\nonumber
\end{align}
and the additional contribution from the boundary layer is%
\begin{align}
\mu_{2}  &  =-2\pi\frac{\sin(aR_{0}+\gamma)}{8\pi a}<K>\\
&  =\frac{1}{4}\left[  \frac{\sin(aR_{0}+\gamma)}{aR_{0}}+\cos(aR_{0}%
+\gamma)\right]  .\nonumber
\end{align}
Combining the mass densities, $\mu=\mu_{1}+\mu_{2}$, yields%
\begin{equation}
4\mu=\cos(\gamma)+\frac{\sin(aR_{0}+\gamma)}{aR_{0}}%
\end{equation}
Substituting from the matching condition, Eq.(\ref{layer-match}),\ the
relation between the linear mass density and the angular deficit is%
\begin{equation}
4\mu-\cos(\gamma)=\pm\delta. \label{m-L}%
\end{equation}
This is not the conventional "thin string" result, $\delta=1-4\mu.$ A possible
explanation is that there is a missing potential energy associated with the
shell assembly, as is found in a matter shell bounding Schwarzschild and
vacuum \cite{Poi04}. However, Futamase and Garfinkle \cite{FG88} have pointed
out that the relation between angular deficit and mass density depends on the
matter in the string, and one could take Eq.(\ref{m-L}) as that relation for
this anisotropic string. A third possibility is that there is additional,
unconsidered structure at the axis. Noting that the metric along$\ r=0$
describes a $2+1$ hypersurface, the $r=0$ axis could be an Israel layer
boundary. This provides additional interior structure whose mass needs to be
considered in the complete calculation.\ With this possibility, the actual
string would be the interior axial structure with the anisotropic
stress-energy described in this example serving as an atmosphere around the string.

\section{Conclusions}

The 3+1 field equations can be written as a set of 2+1 equations on the space
orthogonal to the Killing trajectories.\ In this paper we have presented a
formalism which generalizes a simple 2+1 action for a spacelike Killing vector
to describe a set of fluids with anisotropic stress. The Lagrangian extension
uses both a function $\mathcal{K},$ and a one-form $s_{a}dx^{a}.$\ The spatial
part of $s_{a}$ describes the stress anisotropy in the plane orthogonal to the
Killing vector. With the Killing vector in the (3) direction, $\mathcal{K}%
$\ relates the energy and stress and can describe the anisotropies between the
(1,2) and (3) planes.\ $s_{a}dx^{a}$ is not varied in the action.\ One
advantage of using a fixed form rather than a field is its use as a modeling
tool, with the anisotropies related to complex motions rather than new
physical fields. The anisotropy is described in terms of stress but can be due
to a number of physical mechanisms, such as fluid shear.\ 

Two applications were considered. An anisotropic generalization of the GH
cosmic string interior described an interior solution with positive density
and anisotropic stress bounded by an Israel layer with the GH equation of
state.\ The Bianchi I example described a family of power-law solutions
containing both focusing and unfocusing expansions. The family of
stress-energies was parameterized by a single constant, with the range of the
constant describing the entire stress range: tension through dust to
pressure.\ With the current strong interest in explaining anomalies in the
CMB, one could consider models with a single anisotropic fluid whose stress
shifts from pressure, through dust to tension and focusing to unfocusing
during a series of expansion eras.\ Because the stress associated with the
Killing symmetry is proportional to the Ricci scalar, the curvature will also
evolve.\ In this model there is no expansion along the direction associated
with the Killing coordinate.\ A completely anisotropic expansion could be
generated by considering a higher dimensional action and associating the
Killing symmetry with a higher dimensional manifold.\ Both applications
considered here underline the potential value of formally considering
anisotropic stresses in relativity; while they form a more complicated
stress-energy description, anisotropic fluids can have dynamic features
leading to simpler matter models than those with equal stress.


\begin{thebibliography}{99}                                                                                               %


\bibitem {HOP08}L. Herrera, J. Ospino and A. Di Prisco, \emph{All static
spherically symmetric anisotropic solutions of Einstein's equations, }Phys.
Rev. D\textbf{ 77}, 027502 (2008).

\bibitem {HSW08}L. Herrera, N.O. Santos and A. Wang, $\emph{Shearing}$
$\emph{Expansion-free}$ $\emph{Spherical}$ $\emph{Anisotropic}$ $\emph{Fluid}$
$\emph{Evolution,}$ Phys. Rev. D\textbf{ 78}, 084026 (2008).

\bibitem {CH08}S. Calogero and J.M. Heinzle, \emph{Dynamics of Bianchi type I
solutions of the Einstein equations with anisotropic matter, }arXiv:gr-qc/0809.1008

\bibitem {ACM08}R. Aldrovandi, R.R. Cuzinatto, and L.G. Medeiros,
\emph{Realistic Equations of State for the Primeval Universe, }Eur. Phys. J. C
\textbf{58}, 483 (2008)

\bibitem {KG08}J.P. Krisch and\ E.N. Glass, $\emph{Thin}$ $\emph{shell}$
$\emph{dynamics}$ $\emph{and}$ $\emph{equations}$ $\emph{of}$ $\emph{state,}$
Phys. Rev. D \textbf{78}, 044003 (2008).

\bibitem {KM08}T. Kovisito and D.F. Mota, \emph{Accelerating Cosmologies with
an Anisotropic Equation of State, }Astrophys. J. \textbf{679,} 1 (2008).

\bibitem {BL74}R.L. Bowers and E.P.T. Liang, \emph{Anisotropic spheres in
general relativity}, Astrophys. J. \textbf{188}, 657 (1974).

\bibitem {CHE+81}M. Cosenza, L. Herrera, M. Esculpi, and L. Witten, \emph{Some
models of anisotropic spheres in general relativity}, J. Math. Phys.
\textbf{22}, 118 (1981).

\bibitem {Bay82}S.S. Bayin, \emph{Anisotropic fluid spheres in general
relativity,} Phys. Rev. D \textbf{26}, 1262 (1982).

\bibitem {Bon92}H. Bondi, \emph{Anisotropic spheres in general relativity},
Mon. Not. Roy. Astr. Soc. \textbf{259}, 365 (1992).

\bibitem {Via08}S. Viaggiu, \emph{Modeling Usual and Unsual Anisotropic
Spheres}, Int. J. Mod. Phys. D\textbf{18,} 275 (2009).

\bibitem {DG02}K. Dev and M.\ Gleiser, \emph{Anisotropic stars: \ exact
solutions}, Gen. Rel. Gravit. \textbf{34}, 1793 (2002).

\bibitem {MH03}M.K. Mak and T. Harko, \emph{Anisotropic Stars in General
Relativity,} Proc. Roy.\ Soc. Lon.\ A \textbf{459}, 393 (2003).

\bibitem {DG04}K. Dev and M. Gleiser, \emph{Anisotropic Stars: \ Exact
Solutions and Stability}, Int. J. Mod. Phys\ \textbf{D13}, 1389 (2004).

\bibitem {OTZ+04}A de Oliveira-Costa, M. Tegmark, M.\ Zaldarriaga, and A.
Hamilton, \emph{Significance of the Largest Scale Fluctuations in WMAP}, Phys.
Rev. D \textbf{69}, 063516 (2004).

\bibitem {RS07}A. Raki\'{c} and D.J. Schwarz, \emph{Correlating anomalies of
the microwave sky}, Phys. Rev. D \textbf{75}, 103002 (2007).

\bibitem {GCP07}A.E. Gumrukcuoglu, C.R. Contaldi, and M. Peloso,
\emph{Inflationary perturbations in anisotropic backgrounds and their imprint
on the CMB, }JCAP \textbf{11,} 005 (2007)\emph{ }

\bibitem {PPU07}T.S. Pereira, C. Pitrou and\ J-P Uzan, \emph{Theory of
cosmological perturbations in an anisotropic universe, }JCAP \textbf{09}, 006,(2007).

\bibitem {PPU08}T.S. Pereira, C. Pitrou, and J-P. Uzan, \emph{Predictions from
an anisotropic inflationary era, }JCAP \textbf{04,} 004 (2008).

\bibitem {DD07}M. Demianski and A.G.Doroshkevich, \emph{Possible extensions of
the standard cosmological model: \ anisotropy, rotation and magnetic field,
}Phys. Rev. D \textbf{75}, 123517 (2007).

\bibitem {MKK+07}D.F. Mota, J.R. Kristiansen, T. Koivisto, and N.E.
Groeneboom, \emph{Constraining Dark Energy Anisotropic Stress}, Mon. Not. Roy.
Astron. Soc. \textbf{382, }793 (2007).

\bibitem {Ehl62}J. Ehlers, in \textit{Les Th\'{e}ories Relativistes de la
Gravitation}, Eds. A. Lichnerowicz, M.A. Tonnelat (Colloques Internationaux
C.N.R.S. No. 91, Paris 1962).

\bibitem {Har68}B.K. Harrison, \emph{New Solutions of the Einstein-Maxwell
Equations from Old, }J. Math. Phys. \textbf{9}, 1744 (1968).

\bibitem {Ger71}R. Geroch, \emph{A Method for Generating New Solutions iof
Einstein's Equations I, II, }J. Math. Phys. \textbf{12}, 918 (1971); J. Math.
Phys. \textbf{13}, 394 (1972). \ 

\bibitem {Ste88}H. Stephani, \emph{Symmetries of Einstein's field equations
with a perfect fluid source as examples of Lie-Backlund symmetries, }J. Math.
Phys. \textbf{29}, 1650 (1988).

\bibitem {GGK97}D. Garfinkle, E.N. Glass, and J.P. Krisch, \emph{Solution
Generating with Perfect Fluids, }Gen. Rel. Gravit. \textbf{29,} 467 (1997).

\bibitem {KG02}J.P. Krisch and E.N. Glass, \emph{Adding twist to anisotropic
fluids, }J. Math. Phys. \textbf{43}, 1509 (2002).

\bibitem {Poi04}E. Poisson, \textit{A Relativist's Toolkit} (Cambridge
University Press, Cambridge 2004) p. 40

\bibitem {KE73}A.R. King and G.F.R. Ellis, \emph{Tilted homogeneous
cosmological models}, Commun. Math. Phys. \textbf{31}, 209 (1973).

\bibitem {Tup81}B.O. Tupper, \emph{The equivalence of electromagnetic fields
and viscous fluids in general relativity}, J. Math. Phys. \textbf{22}, 2666 (1981).

\bibitem {KSV08}K. Kuijken, X. Siemens, and T. Vachaspati, \emph{Microlensing
by Cosmic Strings, }Mon. Not. Roy. Astr. Soc. \textbf{384}, 161\ (2008).

\bibitem {FRS+08}A.A. Fraisse, C. Ringeval, D.N. Spergel, and F.R. Bouchet,
\emph{Small-Angle CMB Temperature Anisotropies Induced by Cosmic Strings,
}Phys. Rev. \textbf{D 78}, 043535 (2008).

\bibitem {KW08}R. Khatri and B.D. Wandelt, \emph{Cosmic (super)string
constraints from 21 cm radiation, }Phys. Rev. Lett. \textbf{100}, 091302
(2008). \emph{ }

\bibitem {SSS+08}O.S. Sazhina, M.V. Sazhin, V.N. Sementsov, M. Capaccioli, G.
Longo, G. Riccio and G. D'Angelo, \emph{CMB Anisotropy Induced by a Moving
Straight Cosmic String, }Proceedings: QUARKS-2008, 15th International Seminar
on High Energy Physics, Sergiev Posad, Russia, 23-29 May, 2008.

\bibitem {Vac08}T. Vachaspati, \emph{Cosmic Sparks from Superconducting
Strings, }Phys. Rev. Lett. \textbf{101,} 141301 (2008).

\bibitem {CMM+08}Y. Cui, S.P. Martin, D.E. Morrissey, and J.D. Wells,
\emph{Cosmic Strings from Supersymmetric Flat Directions, }Phys. Rev. D
\textbf{77}, 043528 (2008).

\bibitem {DFM+07}V. Dzhunushaliev, V. Folomeev, K. Myrzakulov. and R.
Myrzakulov, \emph{Cosmic String with two interacting scalar fields, }Mod.
Phys. Lett. A \textbf{22, }407 (2007).

\bibitem {Got85}J.R. Gott III, \emph{Gravitational Lensing Effects of Vacuum
Strings: Exact Solutions, }Astrophys. J \textbf{288}, 422 (1985).

\bibitem {His85}W.A. Hiscock, \emph{Exact gravitational field of a string,
}Phys. Rev. \textbf{31}, 3288 (1985).

\bibitem {FG88}T. Futamase and D. Garfinkle, \emph{What is the relation
between }$\Delta\phi$\emph{ and }$\mu$\emph{ for a cosmic string?}, Phys. Rev.
D \textbf{37}, 2086 (1988).
\end{thebibliography}
\end{document}